\documentclass[prb,twocolumn,amsmath,amssymb,superscriptaddress]{revtex4-2}

\usepackage[pdftex]{graphics}
\usepackage{graphicx}
\usepackage{color}

\begin{document}

\title{RIXS interferometry and the role of disorder in the quantum magnet Ba$_3$Ti$_{3-x}$Ir$_{x}$O$_9$}

\author{M. Magnaterra}
\affiliation{Institute of Physics II, University of Cologne, 50937 Cologne, Germany}
\author{M. Moretti Sala}
\affiliation{ESRF, The European Synchrotron, 71 Avenue des Martyrs, CS40220, 38043 Grenoble Cedex 9, France} 
\affiliation{Dipartimento di Fisica, Politecnico di Milano, I-20133 Milano, Italy}
\author{G. Monaco}
\affiliation{Dipartimento di Fisica e Astronomia "Galileo Galilei", Universit\`{a} di Padova, I-35121 Padova, Italy}
\author{P. Becker}
\affiliation{\mbox{Sect.\ Crystallography, Institute of Geology and Mineralogy, University of Cologne, 50674 Cologne, Germany}}
\author{M. Hermanns}
\affiliation{Department of Physics, Stockholm University, AlbaNova University Center, SE-106 91 Stockholm, Sweden}
\affiliation{Nordita, KTH Royal Institute of Technology and Stockholm University, SE-106 91 Stockholm, Sweden}
\author{P.~Warzanowski}
\author{T. Lorenz}
\author{D.I. Khomskii}
\author{P.H.M. van Loosdrecht}
\affiliation{Institute of Physics II, University of Cologne, 50937 Cologne, Germany}
\author{J. van den Brink}
\affiliation{Institute for Theoretical Solid State Physics, IFW Dresden, 01069 Dresden, Germany}
\affiliation{Institute for Theoretical Physics and W\"urzburg-Dresden Cluster of Excellence ct.qmat, Technische Universit\"at Dresden, 01069 Dresden, Germany}
\author{M. Gr\"{u}ninger}
\affiliation{Institute of Physics II, University of Cologne, 50937 Cologne, Germany}

\date{November 24, 2022}

\begin{abstract}
Motivated by several claims of spin-orbit driven spin-liquid physics in hexagonal 
Ba$_3$Ti$_{3-x}$Ir$_x$O$_9$ hosting Ir$_2$O$_9$ dimers, we report on resonant inelastic 
x-ray scattering (RIXS) at the Ir $L_3$ edge for different $x$. We demonstrate that 
magnetism in Ba$_3$Ti$_{3-x}$Ir$_x$O$_9$ is governed by an unconventional realization 
of strong disorder, where cation disorder affects the character of the local moments.  
RIXS interferometry, studying the RIXS intensity over a broad range 
of transferred momentum ${\mathbf q}$, is ideally suited to assign different 
excitations to different Ir sites. We find pronounced Ir-Ti site mixing. Both ions 
are distributed over two crystallographically inequivalent sites, giving rise to 
a coexistence of quasimolecular singlet states on Ir$_2$O$_9$ dimers and 
spin-orbit entangled $j$\,=\,1/2 moments of $5d^5$ Ir$^{4+}$ ions. 
RIXS reveals different kinds of strong magnetic couplings for different 
bonding geometries, highlighting the role of cation disorder for the 
suppression of long-range magnetic order in this family of compounds.
\end{abstract}

\maketitle

\section{Introduction}

The precise theoretical definition of a quantum spin liquid has been sharpened in 
recent years \cite{Knolle18}, but the elusiveness of experimental realizations has 
remained \cite{Shimizu03,Yamashita10,Han12}. 
Accordingly, the groundbreaking suggestion of Jackeli and Khaliullin \cite{Jackeli09} 
to realize Kitaev's exact spin-liquid solution in, e.g., honeycomb iridates with 
edge-sharing IrO$_6$ octahedra provoked intense activity \cite{WitczakKrempa14,Rau16,Schaffer16,Winter17,Hermanns18,Cao18,Takagi19,Motome20,Takayama21,Trebst22}. 
Dominant bond-directional Kitaev-type exchange interactions indeed were observed in 
Na$_2$IrO$_3$ \cite{Chun15}, but long-range magnetic order arises due to the existence of further 
exchange couplings \cite{Singh10,Singh12,Katukuri14,Rau14,Winter16}. 
In contrast, H$_3$LiIr$_2$O$_6$ was proposed to be a close realization of the Kitaev model, 
evading magnetic order at least down to 50\,mK \cite{Kitagawa18}. However, quantum chemistry 
calculations and density functional theory find a strong sensitivity of exchange couplings 
to H ion disorder and unintentional deuteration \cite{Yadav18,Li18,Wang18}, 
suggesting a more conventional source of magnetic disorder.

\begin{figure}[t]
\centering
\includegraphics[width=\columnwidth]{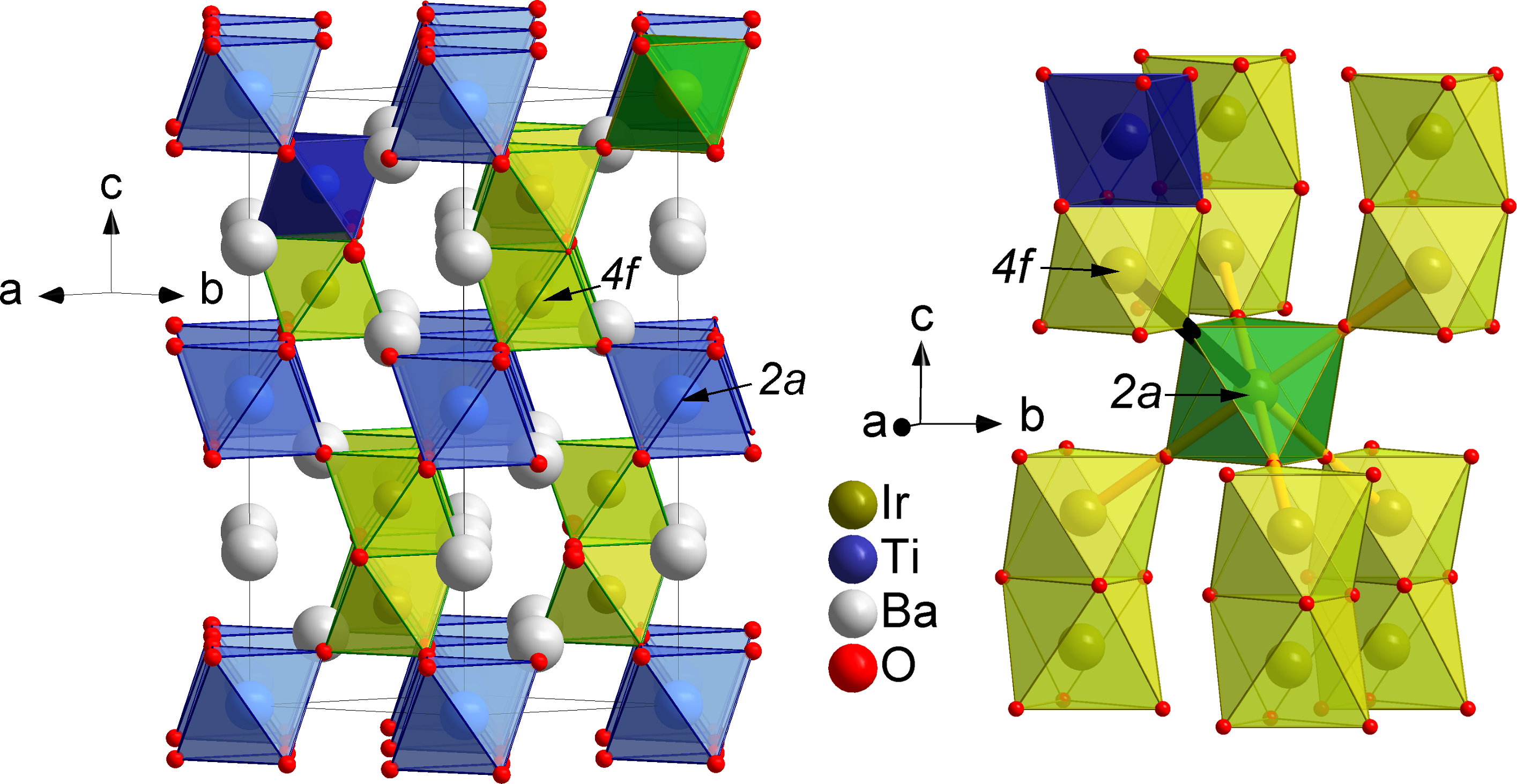}
\caption{Left: Hexagonal crystal structure of Ba$_{3}$Ti$_{3-x}$Ir$_{x}$O$_{9}$. 
	In the hypothetical absence of Ir-Ti site mixing for $x$\,=\,2, Ti (Ir) ions 
	occupy the sites with Wyckoff position \textit{2a} (\textit{4f}). 
	Hence layers of TiO$_6$ octahedra (light blue) are sandwiched between layers of 
	Ir$_2$O$_9$ dimers (light green), the latter being built from face-sharing IrO$_6$ octahedra. 
	The real compound shows Ir-Ti site mixing, i.e., Ir ions also occupy \textit{2a} sites 
	(dark green) and Ti ions also can be found on \textit{4f} sites (dark blue), 
	the latter giving rise to TiIrO$_9$ units instead of dimers. 
	Right: Local structure around an Ir defect ion on a \textit{2a} site (dark green). 
	It is connected to six neighbors on \textit{4f} sites in corner-sharing geometry with 
	$180^\circ$ bonds (thick lines). 
	In this example, five of the six neighbors belong to Ir$_2$O$_9$ dimers 
	but there is a single Ir$^{2a}$-Ir$^{4f}$ pair (thick black line) where the Ir ion 
	on the $4f$ site is part of an IrTiO$_9$ unit.
}
	\label{fig:structure}
\end{figure}

Remarkably, several members of the hexagonal iridate family Ba$_3$$M$Ir$_{2}$O$_9$ were claimed to be 
candidates for spin-liquid behavior \cite{Dey12,Kumar16,Lee17,Dey17,Nag16,Nag19,Khan19,Kumar21}. 
The structure hosts triangular layers of Ir$_2$O$_9$ dimers built from face-sharing 
IrO$_6$ octahedra, see Fig.\ \ref{fig:structure}. 
The versatility of these compounds stems from the valence of the $M$ ions, 
which extends from +1 to +4 
\cite{Doi04,Sakamoto06,Nguyen21,Dey12,Kumar16,Lee17,Dey17,Nag16,Nag19,Kim04,Dey14,Khan19,Kumar21,Garg21}.  
With two holes per dimer such as in Ba$_3$CeIr$_2$O$_9$ with Ce$^{4+}$ and Ir$^{4+}$ ions, 
the ground state is nonmagnetic \cite{Doi04,Streltsov16}. 
RIXS finds quasimolecular orbitals localized on the dimers, i.e., 
the holes are fully delocalized over a given dimer, 
and the tightly bound ground-state singlet forms in a bonding orbital built from 
local $j$\,=\,1/2 moments \cite{Revelli19}. 
In Ba$_3$InIr$_2$O$_9$ with three holes per dimer, persistent spin dynamics were reported 
down to 20\,mK in thermodynamic data and with local probes \cite{Dey17}. In RIXS, 
a quasimolecular $j_{\rm dim}$\,=\,3/2 nature of the local dimer ground state was observed, 
establishing the compound as a cluster Mott insulator \cite{Revelli22}. 
Such spin-orbit-entangled quasimolecular moments with yet to be explored exchange 
interactions may open up a novel route to quantum magnetism \cite{Khomskii21}. 
For four holes per dimer with Ir$^{5+}$ ions, a spin-orbital liquid with weak moments 
has been claimed in Ba$_3$ZnIr$_2$O$_9$ \cite{Nag16,Nag19}. This contradicts the nonmagnetic 
behavior expected for interacting Ir$^{5+}$ $5d^4$ $j$\,=\,0 states and again points 
to a quasimolecular dimer character of the electronic structure.  

In Ba$_3$Ti$_{3-x}$Ir$_x$O$_9$, the comparable ionic radii of Ir$^{4+}$ and Ti$^{4+}$ yield 
Ti-Ir site disorder \cite{Sakamoto06}, see Fig.\ \ref{fig:structure}, and allow for Ir contents 
different from $x$\,=\,2. 
This compound has been discussed as a spin-liquid candidate for $x$\,$\in$\,\{0.5, 1, 2\} 
\cite{Dey12,Kumar16,Lee17}, and the average magnetic moment was reported to \textit{decrease} 
upon increasing the concentration of magnetic Ir$^{4+}$ ions \cite{Kumar16}. 
For $x$\,=\,1, a triangular lattice of Ir$^{4+}$ $j$\,=\,1/2 moments would be formed 
in the hypothetical case of perfect cation order with all Ti ions occupying dimer sites. 
This is attractive from the point of view of theory since novel quantum phases are predicted 
for a triangular Heisenberg-Kitaev model \cite{Becker15,Catuneanu15}.
To understand the magnetism of Ba$_3$Ti$_{3-x}$Ir$_x$O$_9$ one needs to address both  
disorder and the role of quasimolecular dimer states. 
We show below that RIXS is the ideal tool to achieve this goal.

In $L$ edge RIXS measurements on Ba$_3$Ti$_{3-x}$Ir$_x$O$_9$ with $x$\,$\in$\,\{0.3, 0.5, 1.5, 1.8\}, 
we find a coexistence of quasimolecular Ir$_2$ dimer states and single-site $j$\,=\,1/2 moments, 
providing a clear fingerprint of substantial Ir-Ti site mixing. 
The quasimolecular dimer character is nailed down via a double-slit-type sinusoidal 
interference pattern in the RIXS intensity as a function of the transferred momentum $\mathbf{q}$, 
in agreement with previous results on Ba$_3$CeIr$_2$O$_9$ and Ba$_3$InIr$_2$O$_9$ \cite{Revelli19,Revelli22}.
Additionally, the study of such interference patterns allows us to identify a RIXS peak 
at 0.15\,eV with site-mixing-induced magnetic excitations of Ir moments on $M$ sites ($2a$) 
that are strongly exchange-coupled to neighboring dimer sites ($4f$) via 180$^\circ$ bonds, 
see Fig.\ \ref{fig:structure}.
The $2a-4f$ bonding geometry is very similar to the case of Sr$_2$IrO$_4$
\cite{Kim12,KimNatComm14} which roughly explains the energy scale of 0.15\,eV.\@ The ability to 
disentangle the contributions of different Ir sites in such a strongly disordered system 
demonstrates the power and versatility of RIXS interferometry.

Our findings on Ba$_3$Ti$_{3-x}$Ir$_x$O$_9$ highlight the unconventional role that disorder may play 
in a cluster Mott insulator. Removing a magnetic ion from a cluster such as a dimer does not 
create a usual vacancy but strongly changes the moment's character as well as the 
relevant (exchange) interactions. 
In our example of dimers, the character changes from quasimolecular $j_{\rm dim}$\,=\,$0$ 
to single-site $j$\,=\,$1/2$ moments. For perfect cation order, magnetic interactions 
for the Ir$_2$ dimers range from very weak for interdimer exchange couplings up to 
about 1\,eV for intradimer singlet-to-triplet excitations, but cation disorder yields 
pairs of local moments with exchange coupling of about 0.15\,eV.\@
The physics of Ba$_3$Ti$_{3-x}$Ir$_{x}$O$_9$ thus is governed by a complex mixture 
of Ti-Ir site disorder, the statistical coexistence of different magnetic moments, 
and very different magnetic couplings.

\section{Experimental}

Crystals of Ba$_3$Ti$_{3-x}$Ir$_{x}$O$_9$ were grown using BaCO$_3$, IrO$_2$, and TiO$_2$ 
as educts and BaCl$_2$ as melt solvent. After the growth process the crystals were 
mechanically separated from the flux and washed with cold H$_2$O.\@ 
We studied samples with an Ir content $x \in\{0.3, 0.5, 1.5, 1.8\}$ as determined by 
energy dispersive x-ray spectroscopy (EDX). The crystals with $x$\,=\,0.3 and 0.5 resulted 
from one growth experiment with a ratio IrO$_2$:TiO$_2$\,=\,1:2, while the crystals with 
$x$\,=\,1.5 and 1.8 were obtained for IrO$_2$:TiO$_2$\,=\,2:1 and otherwise identical 
growth parameters. 
This indicates an effective distribution coefficient 
$k_{\rm eff}$\,=\,$c_{\rm Ir}^{\rm cryst}/c_{\rm Ir}^{\rm melt}  < 1$ for the concentration 
of Ir ions in the crystal and the melt, respectively. This causes Ir depletion in the 
crystals with respect to the melt and thus Ir accumulation in the melt as a function of time. 
Crystals nucleating later in the course of the unseeded growth experiment 
grow in a melt that is enriched with Ir, giving rise to a larger Ir content $x$. 
From single crystal X-ray diffraction and structure refinements, 
room-temperature lattice constants of $a$\,=\,$(5.7103 \pm 0.0011)$\,\AA\ and 
$c$\,=\,$(14.1516 \pm 0.0029)$\,\AA\ were obtained 
in the hexagonal space group $P6_{3}/mmc$ for two samples from the two different batches. 
The magnetic susceptibility was measured in a commercial SQUID magnetometer (Quantum Design MPMS).

RIXS measurements were performed in horizontal scattering geometry at the Ir $L_{3}$ edge 
at beamline ID20 at the ESRF \cite{Moretti13,Moretti18}. We measured on the (001) surface 
for $x$\,=\,0.3 and $x$\,=\,1.8 and on the (100) surface for $x$\,=\,0.5 and $x$\,=\,1.5. 
To determine the resonance behavior, we collected a low-resolution (0.4\,eV) RIXS map at 
300\,K on Ba$_3$Ti$_{1.2}$Ir$_{1.8}$O$_9$, see Sec.\ \ref{sec:RIXS}. 
Focusing on intra-$t_{2g}$ excitations, we choose an incident energy of 11.215 keV.\@
At $T$\,=\,20\,K, we measured RIXS spectra with an energy resolution of 27\,meV by scanning 
the energy loss at constant {\bf q}. The RIXS spectra were normalized with respect to the 
incident flux and the acquisition time. Furthermore, we collected {\bf q} scans at constant 
energy loss. The intensity was corrected for self-absorption effects, 
compensating the different paths traveled by the x-ray photons within the absorbing sample 
for different scattering geometries \cite{Minola15}. 

The two sites of a dimer are connected by the vector $\mathbf{d}$\,=\,(0,\,0,\,$d$). 
As discussed below, the quasimolecular dimer features are characterized by a RIXS intensity 
that is modulated with a period $2Q_d$\,=\,$2\pi/d$ which is incommensurate with the 
Brillouin zone. Accordingly, we denote reciprocal space vectors in absolute units 
while the indices $(h\,\,k\,\,l)$ are given in reciprocal lattice units (r.l.u.).

\section{Cation disorder}

In the absence of cation disorder, the $M$ ions in Ba$_3$$M$Ir$_2$O$_9$ occupy the 
single-layer $2a$ sites while the Ir ions are located on the dimer $4f$ sites, 
see Fig.\ \ref{fig:structure}. 
In Ba$_3$$M$Ir$_2$O$_9$, perfect cation ordering is achieved for a significant difference 
in ionic radii of $M^{4+}$ and Ir$^{4+}$ \cite{Sakamoto06,Revelli19}. 
In contrast, site mixing of Ti$^{4+}$ and Ir$^{4+}$ has been observed in 
Ba$_3$Ti$_{3-x}$Ir$_x$O$_9$ \cite{Sakamoto06}, which can be rationalized by the fact that 
Ti$^{4+}$ and Ir$^{4+}$ do not only show the same valence but also similar ionic radii. 
On the $2a$ position, the Ir-Ti site mixing in Ba$_3$Ti$_{3-x}$Ir$_x$O$_9$ was reported 
to amount to about 6\,\% and 21\,\% for $x$\,=\,1 and 2, 
respectively \cite{Dey12,Kumar16,Lee17,Sakamoto06}, suggesting that Ir preferably occupies 
$4f$ dimer sites. Within the dimers, there is no preferential occupation of Ir$^{4+}$ 
and Ti$^{4+}$ between the two crystallographically equivalent dimer 
sites \cite{Dey12,Kumar16,Lee17,Sakamoto06}.

\section{Magnetic susceptibility}
\label{sec:susc}

\begin{figure}[t]
\centering
\includegraphics[width=\columnwidth]{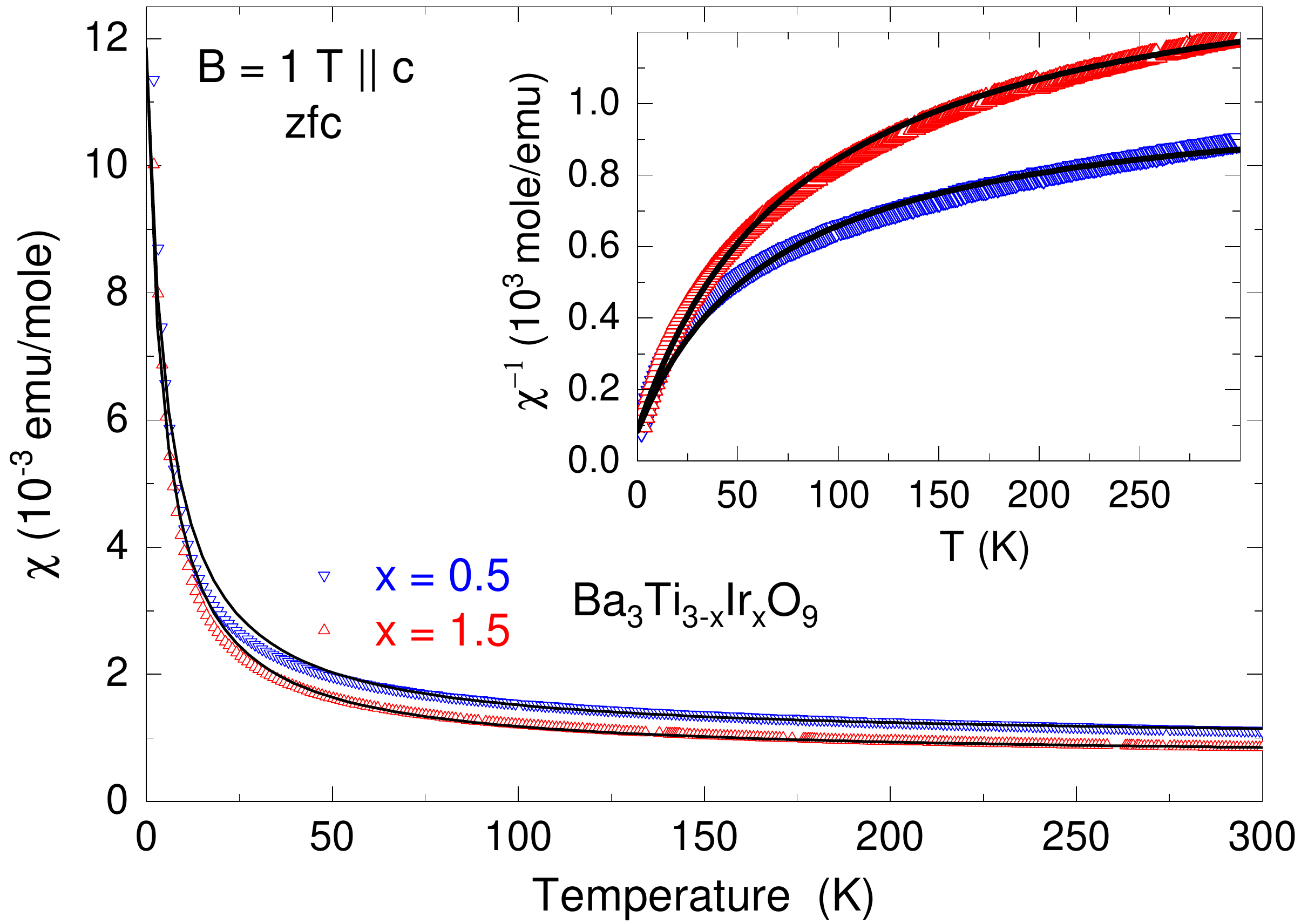}
\caption{Magnetic susceptibility $\chi(T)$ of Ba$_3$Ti$_{3-x}$Ir$_{x}$O$_9$ for $x$\,=\,0.5 
	and 1.5, measured on the same samples as studied in RIXS.\@ 
	The inset displays $1/\chi$ versus $T$. Solid lines: Curie-Weiss fits according 
	to Eq.\ (\ref{eq:CW}) with fixed $\mu_{\rm eff}$\,=\,$1.73$\,$\mu_{\rm B}$.
}
\label{fig:chi}
\end{figure}

The magnetic susceptibility $\chi(T)$ is plotted in Fig.\ \ref{fig:chi} for two samples 
with $x$\,=\,0.5 and 1.5. 
After zero-field cooling (zfc), the data were obtained in a field of 1\,T upon heating, 
but there is no difference to the corresponding data obtained after field cooling in 
1\,T (not shown). Our data do not provide any indication for long-range magnetic order and the 
overall magnetic signal does not increase from $x$\,=\,0.5 to 1.5, i.e., upon enhancing 
the content $x$ of magnetic Ir$^{4+}$ ions by a factor of three. The latter at first sight 
is surprising but it agrees with previous results \cite{Kumar16}.  
This unusual behavior can be explained using our RIXS results, see Sec.\ \ref{sec:RIXS}, 
which demonstrate strong site disorder of Ir$^{4+}$ and Ti$^{4+}$ ions with respect to 
the occupation of the $(2a)$ $M$ sites and the $(4f)$ dimer sites, 
see Fig.\ \ref{fig:structure}, in combination with the formation of nonmagnetic singlets 
in the case of two Ir$^{4+}$ ions occupying both $4f$ sites within a dimer. 
Therefore, the overall magnetic susceptibility consists of the superposition of different 
Curie-Weiss and van Vleck contributions based on local moments on $2a$ or $4f$ sites and dimers. 
Because a reliable quantitative separation of the various contributions 
to $\chi(T)$ is not possible, we restrict ourselves to a strongly simplified effective 
Curie-Weiss model, which neglects a possible temperature dependence of the effective 
magnetic moments by setting 
$\mu_{\rm eff}$\,=\,1.73\,$\mu_{\rm B}$ of a spin 1/2. 
Moreover, we assume temperature-independent van Vleck terms which are included in a 
constant background susceptibility $\chi_0$, 
\begin{equation}\label{eq:CW}
	\chi(T)= n  \frac{N_{\rm A}\, \mu_{\rm eff}^2}{3k_{\rm B}(T-\theta_W)}+\chi_0 \, .
\end{equation}
This yields three adjustable parameters, where $n$ denotes the amount of localized 
magnetic moments, $\theta_W$ measures their effective interaction strength,  
and $\chi_0$ is the sum of the overall van Vleck contribution and the core diamagnetism. 
As shown in Fig.\ \ref{fig:chi}, the measured data can be reasonably well described by 
the simplified model of Eq.\ (\ref{eq:CW}) with the parameter set $n$\,=\,0.16, 
$\theta_W$\,=\,$-5.5$\,K, and $\chi_0$\,=\,$9.5\cdot 10^{-4}$emu/mole for $x$\,=\,0.5, 
while we obtain $n$\,=\,$0.14$, $\theta_W$\,=\,$-4.7$\,K, and 
$\chi_0$\,=\,$6.8\cdot 10^{-4}$emu/mole for the sample with $x$\,=\,1.5. 
Both parameter sets with weakly antiferromagnetic Weiss temperatures and comparable 
$\chi_0$ values are very similar, as expected due to the very similar $\chi(T)$ data. 
In view of the very different values of $x$, the very similar results 
for $n$ are remarkable, but our RIXS data confirm a similar density of local moments 
in the two samples, see Sec.\ \ref{sec:single}. This indicates that for large $x$ 
the vast majority of Ir$^{4+}$ ions form singlet dimers.

\begin{figure}[t]
\centering
\includegraphics[width=1.0\columnwidth]{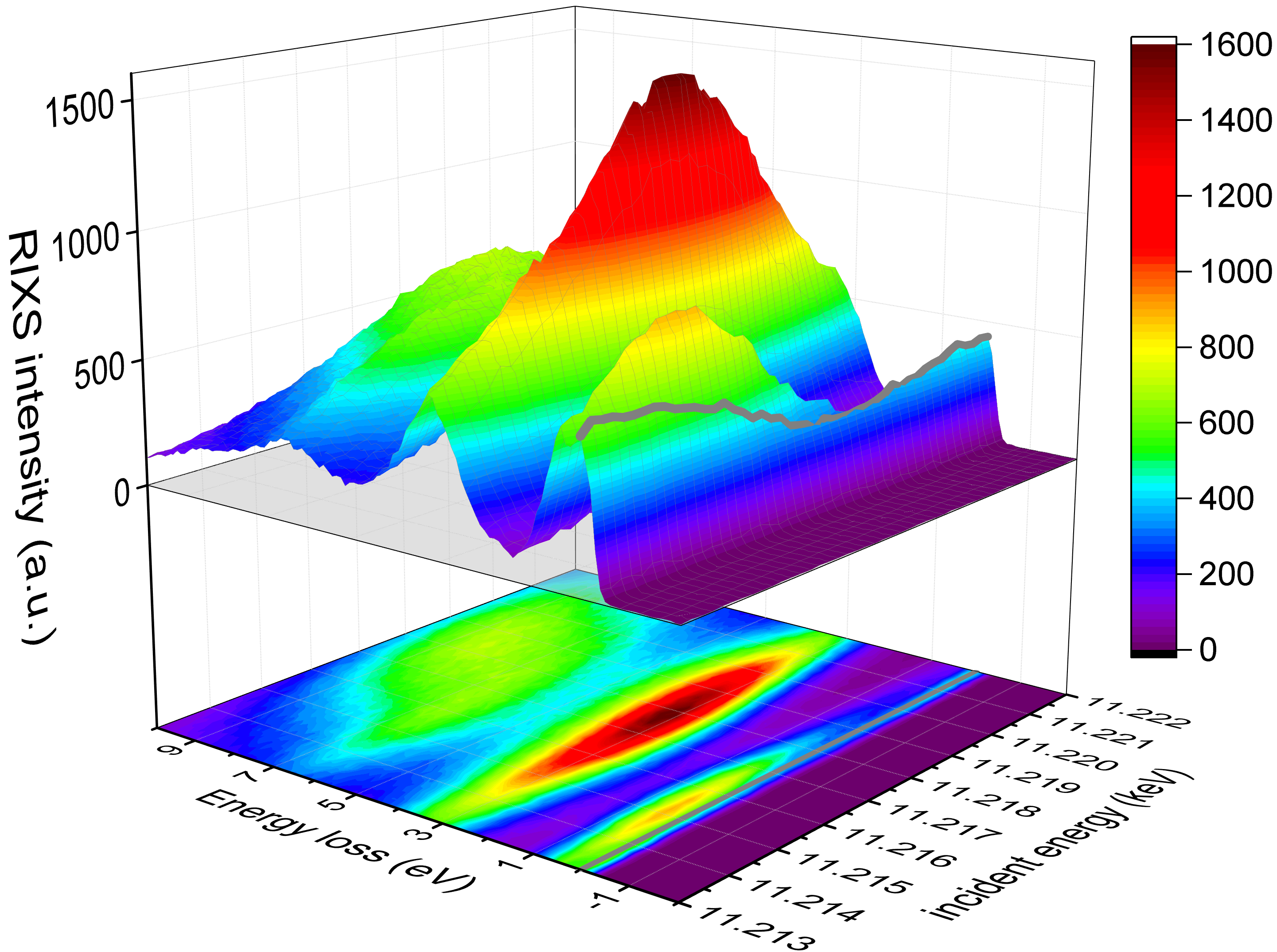}
\caption{RIXS resonance map of Ba$_{3}$Ti$_{1.2}$Ir$_{1.8}$O$_{9}$  measured at 300\,K 
	for transferred momentum $\mathbf{q}$\,=\,(0\,\,0\,\,18.5) r.l.u..
	The dominant RIXS peak at 3\,eV energy loss corresponds to excitations to $e_g^\sigma$ orbitals which are resonantly enhanced for an incident energy of about 11.218\,keV.\@ 
	The intra-$t_{2g}$ excitations below about 1.2\,eV are maximized for an incident energy 
	of 11.215\,keV.\@  
	Thick grey line: elastic line at zero loss.}
\label{fig:map}
\end{figure}

\section{RIXS results}
\label{sec:RIXS}

\begin{figure}[t]
\centering
\includegraphics[width=\columnwidth]{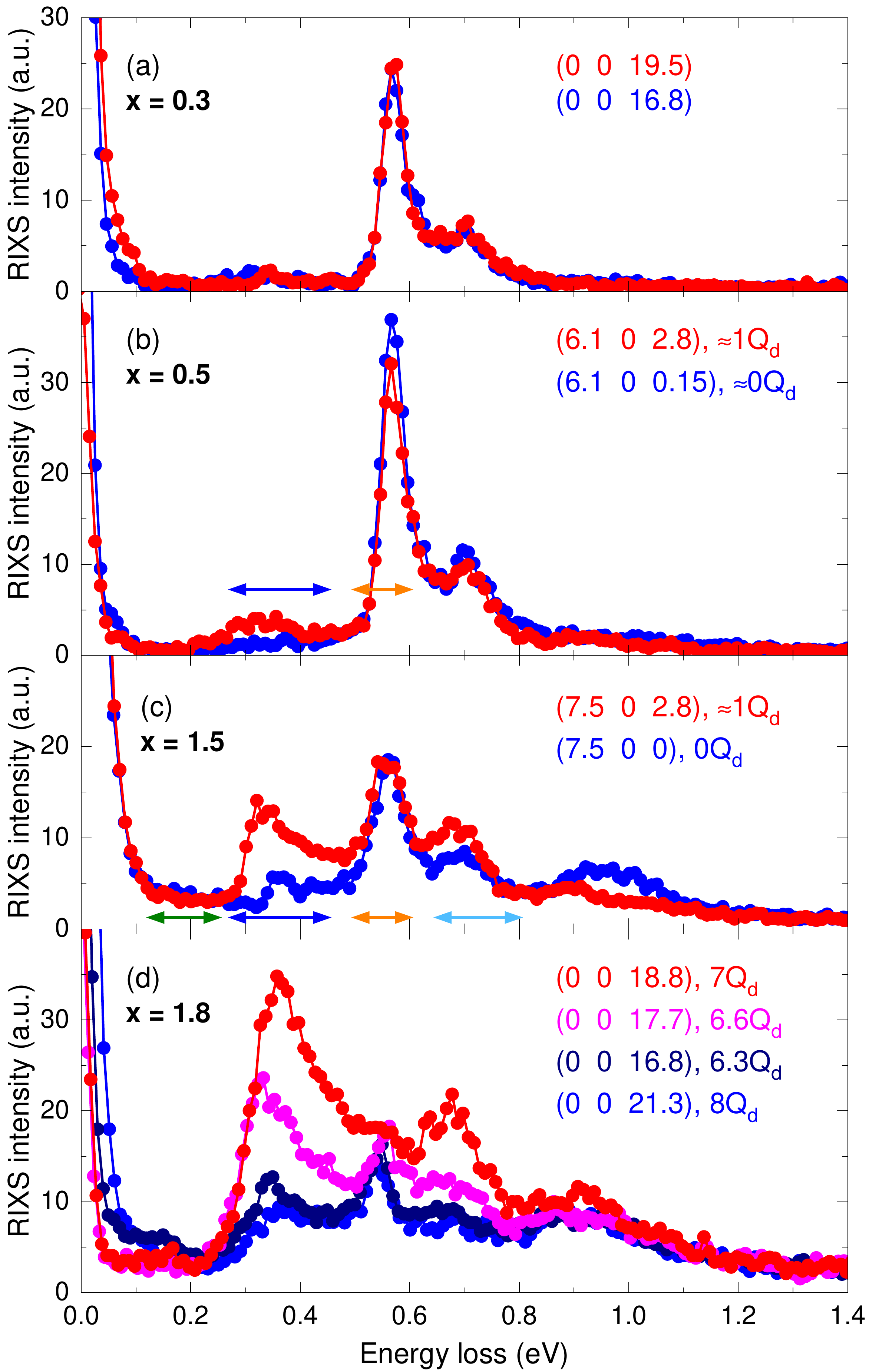}
\caption{RIXS spectra of Ba$_{3}$Ti$_{3-x}$Ir$_{x}$O$_{9}$ at $T$\,=\,20\,K, showing 
	the coexistence of single-site and dimer excitations at least for $x \geq 0.5$. 
	For each $x$, we compare data for different $l$. 
	For $x$\,=\,0.3 and 1.8 we measured on the (001) surface with $h$\,=\,$k$\,=\,0, 
	while the data for $x$\,=\,0.5 and 1.5 were collected on the (100) surface 
	with large $h$ and smaller $l$. 
	For small $x$, the two-peak structure of the spin-orbit exciton dominates, 
	a mark of single-site $j$\,=\,1/2 moments.  
	For large $x$, the broad features above 0.2\,eV correspond to 
	quasimolecular excitations on Ir$_2$ dimers, which is reflected in the 
	periodic $q_c$ dependence of the intensity, cf.\ Fig.\ \ref{fig:l_scan_Ir05}. 
	The arrows in panels (b) and (c) denote the integration ranges 
	used in Fig.\ \ref{fig:l_scan_Ir05}.
}
\label{fig:4samples}
\end{figure}

\begin{figure}[t]
\centering
\includegraphics[width=\columnwidth]{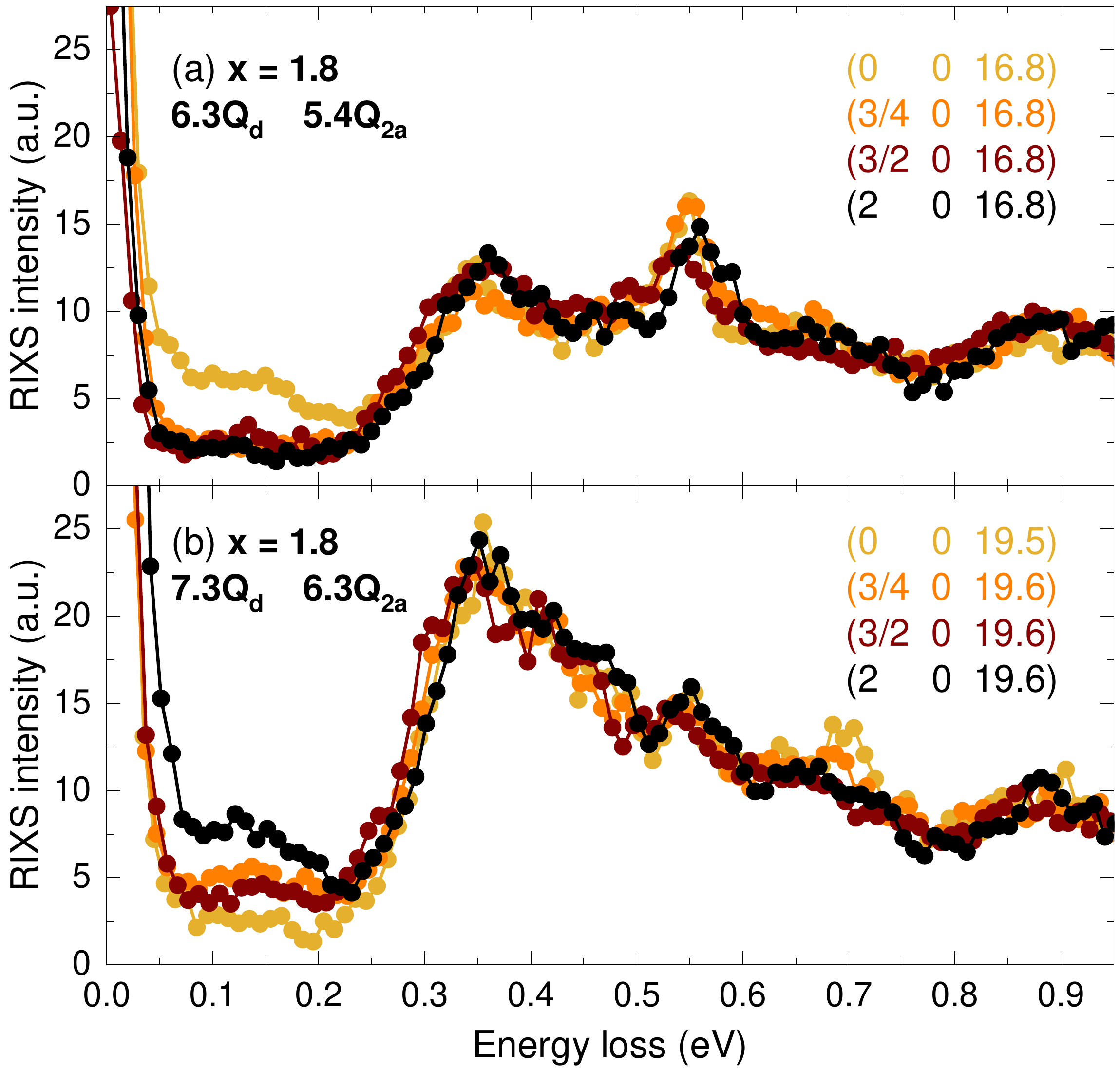}
\caption{RIXS spectra of Ba$_{3}$Ti$_{1.2}$Ir$_{1.8}$O$_{9}$ for $0\leq h \leq 2$ 
	at $T$\,=\,20\,K.\@ 
	For the dimer excitations above 0.2\,eV, the different intensities in the top and 
	bottom panels reflect the $\sin^2(q_c d/2)$ modulation. With $Q_d$\,=\,$\pi/d$, 
	the value $l$\,=\,16.8 (19.6) corresponds to $q_c$\,$\approx$\,$6.3Q_d$ ($7.3Q_d$) 
	which is close to a minimum (maximum), see Figs.\ \ref{fig:4samples} and 
	\ref{fig:l_scan_18}. 
	Only the weak 0.15\,eV peak exhibits a pronounced $h$ dependence. It corresponds to a 
	magnetic excitation that involves an Ir ion on a $2a$ site, giving rise to an 
	intensity modulation along $h$, $k$, and $l$, see Sec.\ \ref{sec:2a}. 
	For constant $h$ and $k$, it shows a period of $2Q_{2a} \approx 2.3 Q_d$. 
	In contrast to the dimer features above 0.2\,eV, its intensity for $h$\,=\,$k$\,=\,0 
	is enhanced for $l$\,=\,16.8 but suppressed for $l$\,=\,19.6. 
	}
	\label{fig:spectra_h}
\end{figure}

A low-resolution (0.4\,eV) RIXS map of Ba$_3$Ti$_{1.2}$Ir$_{1.8}$O$_9$ illustrates the 
resonance enhancement, see Fig.\ \ref{fig:map}. 
We find dominant excitations into $e_g^\sigma$ levels above about 3\,eV energy loss 
and charge-transfer excitations at still higher energies. In the following, we focus 
on the intra-$t_{2g}$ excitations below about 1.2\,eV.\@ 
RIXS spectra of Ba$_3$Ti$_{3-x}$Ir$_x$O$_9$ for $x$\,$\in$\,\{0.3, 0.5, 1.5, 1.8\} are 
depicted in Fig.\ \ref{fig:4samples} for selected values of $\mathbf{q}$. 
Each panel shows data for different $q_c$, the component of $\mathbf{q}$ 
parallel to the $c$ axis. 
To identify dimer features (see below), we express $q_c$ in terms of 
$Q_d$\,=\,$\pi/d \approx 5.3 \times \pi/c$. 
In contrast, Fig.\ \ref{fig:spectra_h} highlights the $h$ dependence 
for $q_c$\,$\approx$\,$6.3Q_d$ and $7.3Q_d$ for $x$\,=\,1.8. 

The data in Fig.\ \ref{fig:4samples} demonstrate the coexistence of individual 
$j$\,=\,1/2 sites and dimers with quasimolecular states. 
The overall line shape changes strongly as a function of $x$ since the contribution 
of $j$\,=\,1/2 sites dominates for small $x$, see Sec.\ \ref{sec:single}, 
while quasimolecular dimer excitations prevail for large $x$. 
The dimer excitations set in above about 0.2\,eV, see Sec.\ \ref{sec:dimer}. 
For $x$\,=\,1.5 and 1.8, we find a further, weak RIXS peak at about 0.15\,eV which is 
most evident in Fig.\ \ref{fig:spectra_h} as the only feature with a particular 
$h$ dependence of the intensity. Using RIXS interferometry we demonstrate that 
this feature corresponds to magnetic excitations related to Ir ions on $2a$ sites, 
see Sec.\ \ref{sec:2a}. 

The elastic line can be suppressed for a scattering angle $2\theta$\,$\approx$\,$90^\circ$ 
for incident $\pi$ polarization. This is illustrated by the data in 
Fig.\ \ref{fig:4samples}(d) that were measured with 
$2\theta$\,$\approx$\,$82^\circ$, $87^\circ$, $94^\circ$, and $112^\circ$
for $l$\,=\,16.8, 17.7, 18.8, and 21.3, respectively.

\subsection{Individual $j$\,=\,1/2 moments}
\label{sec:single}

For small $x$, the RIXS spectra provide an unmistakable fingerprint of diluted 
$5d^5$ Ir$^{4+}$ sites. The data predominantly show the well-known two-peak structure 
of the so-called spin-orbit exciton, i.e., excitations from local $j$\,=\,1/2 moments
to the $j$\,=\,3/2 excited quartet state which is, however, split by the trigonal 
crystal field \cite{Moretti14}. 
Features similar to the two narrow peaks at 0.57 and 0.70\,eV were observed in many 
iridates with (weakly) interacting $j$\,=\,1/2 
moments \cite{Gretarsson13,Liu12,Rossi17,Revelli19b,Ruiz21,delaTorre21,Reig20,Khan21}. 
Compared to, e.g., honeycomb Na$_2$IrO$_3$ or square-lattice Sr$_2$IrO$_4$ \cite{Gretarsson13,Kim12,KimNatComm14}, 
the small line width of the pronounced peak at 0.57\,eV underlines that interactions 
between individual $j$\,=\,1/2 moments are weak in Ba$_3$Ti$_{3-x}$Ir$_x$O$_9$.

Typical for iridates, Ba$_3$Ti$_{3-x}$Ir$_{x}$O$_9$ shows a strong cubic crystal-field 
splitting 10\,$Dq$\,$\approx$\,3\,eV between $t_{2g}$ and $e_g^\sigma$ states, 
see Fig.\ \ref{fig:map}. In this case, the physics of intra-$t_{2g}$ excitations 
on a single $t_{2g}^5$ site is well described by the Hamiltonian \cite{Moretti14}
\begin{equation}
H_{\rm single} = \lambda\, \mathbf{S}\cdot \mathbf{L} + \Delta_{\rm CF} L_z^2 \, , 
\label{eq:single}
\end{equation}
where $\lambda$ denotes spin-orbit coupling and $\Delta_{\rm CF}$ the trigonal 
crystal field. In terms of an IrO$_6$ point-charge model, trigonal elongation or 
contraction of the oxygen octahedron corresponds to positive or negative $\Delta_{\rm CF}$, 
but in Ba$_3$Ti$_{3-x}$Ir$_x$O$_9$ with face-sharing octahedra the sign also depends 
on covalency and interactions with further ions \cite{KhomskiiZhETF}. 
Based on the crystal structure, $\Delta_{\rm CF}$\,$\neq$\,$0$ is already expected 
for undistorted octahedra. Depending on the sign of $\Delta_{\rm CF}$, the observed 
peak energies of 0.57\,eV and 0.70\,eV yield either 
$\lambda$\,=\,0.41\,eV and $\Delta_{\rm CF}$\,=\,-0.18\,eV or 
$\lambda$\,=\,0.40\,eV and $\Delta_{\rm CF}$\,=\,0.23\,eV.\@ 
Since the Hamiltonian in Eq.\ (\ref{eq:single}) only deals with $t_{2g}$ electrons, it yields 
an effective value of $\lambda$ that slightly depends on $10\,Dq$. 
Our solutions $\lambda$\,=\,0.40\,eV or 0.41\,eV fall well within the range reported in RIXS 
on Ir oxides 
\cite{Gretarsson13,Liu12,Revelli19b,Ruiz21,delaTorre21,Kim12,KimNatComm14,Vale19,Yuan17,Kusch18,Nag18,Wang19,Katukuri22,Revelli19,Revelli22}.

With Ir$^{4+}$ ions occupying $4f$ and $2a$ sites, one may expect different values 
of $\Delta_{\rm CF}$, and these may cause a further splitting of the peaks at 0.57\,eV 
and 0.70\,eV.\@ As stated above, only about 6\,\% of the $2a$ sites were reported to be occupied 
by Ir ions for $x$\,=\,1 but this value increases to 21\,\% 
for $x$\,=\,2 \cite{Dey12,Kumar16,Lee17,Sakamoto06}. 
For large $x$, a detailed analysis of the spin-orbit exciton peaks is hindered by the 
additional presence of dimer features in the spectra. However, the enhanced width of 
the 0.57\,eV peak for $x$\,=\,1.5 possibly reflects different values of $\Delta_{\rm CF}$ 
on the two sites.

Comparing $x$\,=\,0.3 and 0.5, the RIXS intensity of the peak at 0.57\,eV increases with $x$, 
see Fig.\ \ref{fig:4samples}, and the relative intensity roughly agrees with the relative 
Ir concentration. 
In the same spirit, the integrated intensity of the 0.57\,eV peak for $x$\,=\,0.15 amounts 
to about 80\,\% of the value for $x$\,=\,0.5, which roughly agrees with the ratio of the 
$n$ values determined from $\chi(T)$, cf.\ Sec.\ \ref{sec:susc}.
However, the RIXS matrix elements depend on the measurement geometry and the 
sample orientation,
and the intensity is further affected by the surface quality. Therefore, we refrain 
from a more detailed quantitative comparison of the RIXS intensities obtained on 
different samples. However, we emphasize that the strong peak at 0.57\,eV is observed 
for all $x$, demonstrating the presence of individual $j$\,=\,1/2 sites even for large $x$ 
when the response is dominated by dimers.

\subsection{Quasimolecular dimer excitations}
\label{sec:dimer}

For $x$\,=\,0.5, at most 1/4 of the $4f$ dimer sites can be occupied by Ir ions, and 
this number is even smaller if Ir ions also occupy the $2a$ sites. 
The pronounced change of the line shape of the RIXS spectra with increasing $x$, 
cf.\ Fig.\ \ref{fig:4samples}, reflects an increasing density of Ir$_2$ dimers 
along with a dramatic difference of the electronic structure of dimers compared 
to weakly interacting $j$\,=\,1/2 moments. 
For $x$\,=\,1.8, the strongest RIXS peak is observed at about 0.35\,eV, and the 
spectra for $x$\,=\,1.5 and 0.5 reveal corresponding features at the same energy, 
i.e., dimers are already formed for $x$\,=\,0.5.

\begin{figure}[t]
\centering
\includegraphics[width=\columnwidth]{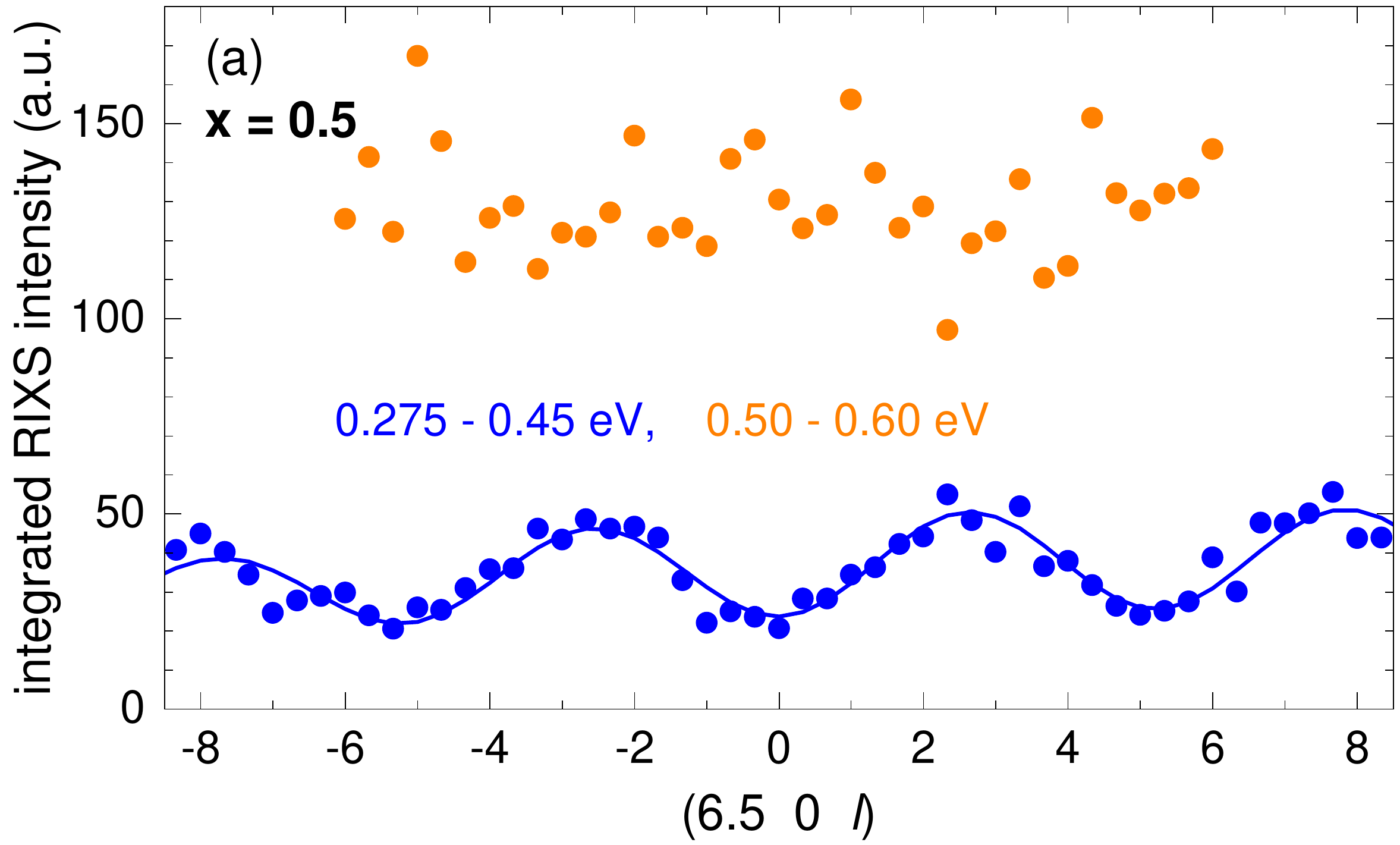}
\includegraphics[width=\columnwidth]{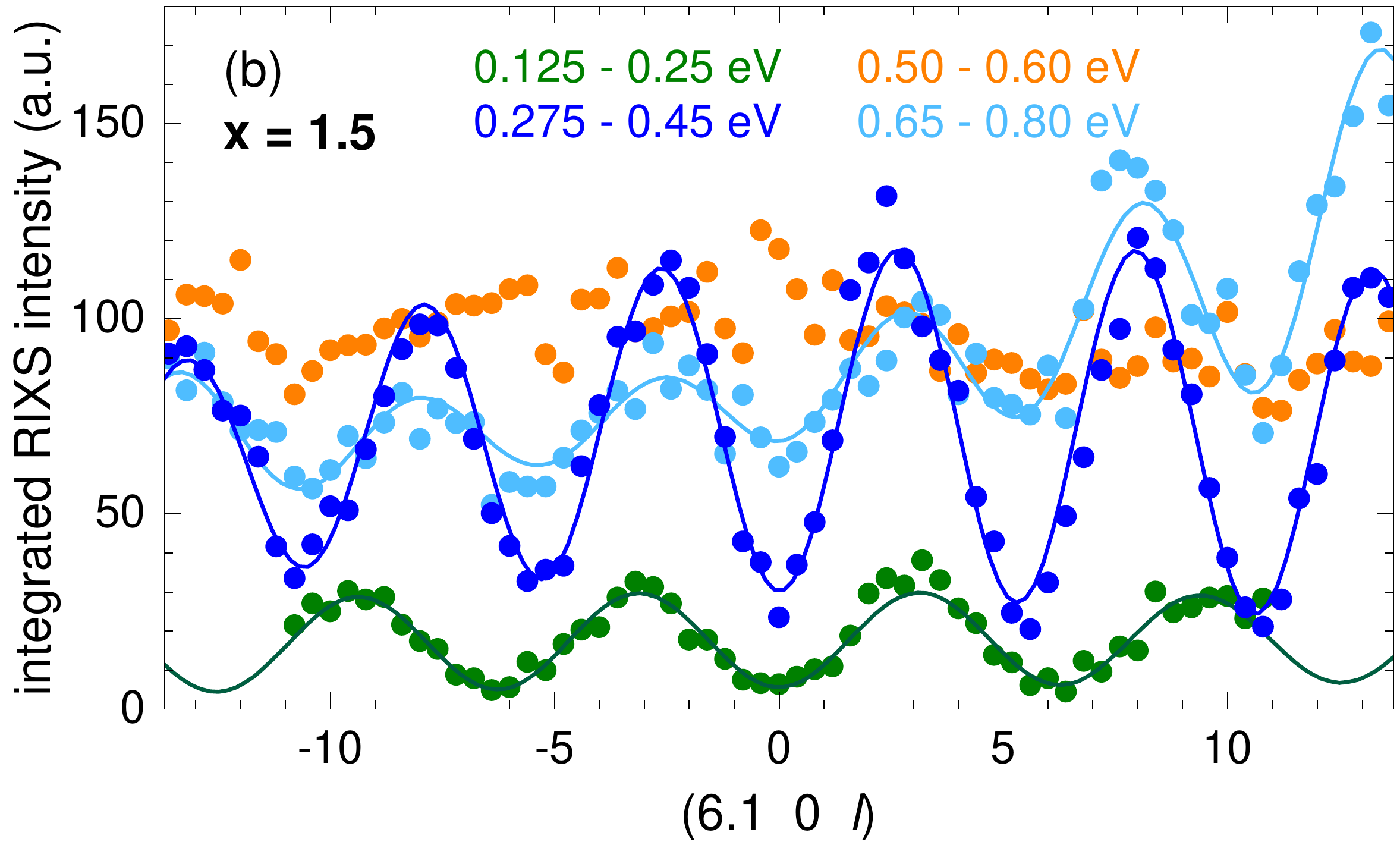}
\caption{RIXS interference patterns as a function of $l$ for $x$\,=\,0.5 (a) 
	and 1.5 (b). Measuring on a $(100)$ surface with $c$ in the scattering plane 
	allows us to cover a large range of $l$ that includes $l$\,=\,0. 
    Data were corrected for self absorption \cite{Minola15} and 
	integrated over the indicated energy ranges. 
	The local, single-site character of the spin-orbit exciton yields roughly 
	constant intensity around 0.57\,eV (orange). 
	In contrast, the $\sin^2(q_c\,d/2)$ modulation of the 0.35\,eV peak (dark blue) 
	with period $2Q_d$\,=\,$2\pi/d$ demonstrate its quasimolecular dimer character. 
	The larger period observed for integration below 0.25\,eV for $x$\,=\,1.5 (green in (b)) 
	unveils a different microscopic origin related to Ir ions on $2a$ sites, 
	see Sec.\ \ref{sec:2a}. For each point, the integration time was 20\,s (orange) 
	and 60\,s (blue) in the top panel and 30\,s in the bottom panel.
}	
\label{fig:l_scan_Ir05}
\end{figure}

In spectroscopy on crystalline materials, studies of the $\mathbf{q}$-dependent 
properties often focus on the dispersion $\omega(\mathbf{q})$ of the excitation energy. 
For excitations localized on a dimer, however, the key feature is the 
$\mathbf{q}$ dependence of the RIXS \textit{intensity}, allowing us to unravel 
the quasimolecular character of a given excitation \cite{Revelli19,Revelli22}. 
For an individual $j$\,=\,1/2 moment on a single site, neglecting interactions, 
the local excitation to $j$\,=\,3/2 states does not show any $\mathbf{q}$ dependence. 
The spin-orbit exciton peaks at 0.57 and 0.70\,eV support this picture for 
$x$\,=\,0.3 and 0.5 for two different $\mathbf{q}$ points, see Fig.\ \ref{fig:4samples}. 
Addressing the $\mathbf{q}$ dependence more explicitly, Fig.\ \ref{fig:l_scan_Ir05} 
shows that the integrated RIXS intensity of the strong 0.57\,eV peak is roughly 
independent of $q_c$ over many Brillouin zones. 
This is observed for both $x$\,=\,0.5 and 1.5. 
In contrast, the sinusoidal modulation of the RIXS intensity integrated around 0.35\,eV 
(dark blue symbols in Fig.\ \ref{fig:l_scan_Ir05}) provides an unambiguous proof of 
the quasimolecular character of the orbitals involved in this excitation \cite{Revelli19}, 
again both for $x$\,=\,0.5 and 1.5.  
The data are well described by
\begin{equation}
I_f(q_c) = (a_0 + a_1 q_c + a_2 q_c^2) \sin^2(q_c \,d/2) + c_0 + c_1 q_c\, ,
\label{eq:Iqc}
\end{equation}
where $q_c$ is the $\mathbf{q}$-component parallel to the $c$ axis, 
$\mathbf{d}$\,=\,$(0,0,d)$ connects two Ir $4f$ sites within a dimer, 
and the $a_i$ and $c_i$ are fit parameters describing the modulation amplitude 
and an offset, respectively. 
The $\mathbf{q}$ dependences of amplitude and offset mainly reflect the change of the 
experimental scattering geometry that is necessary to vary $\mathbf{q}$ over a 
large range, giving rise to a change of polarization factors. At the same time, 
the scattering geometry determines self-absorption effects \cite{Minola15} which, 
however, have been corrected in our data.

For $x$\,=\,1.5 and integration from 0.65 to 0.80\,eV, we find a modulation 
with the same period as around 0.35\,eV but reduced amplitude (light blue symbols 
in Fig.\ \ref{fig:l_scan_Ir05}), 
suggesting that dimer excitations overlap with the single-site 0.70\,eV peak. 
Remarkably, integration below 0.25\,eV (green symbols in Fig.\ \ref{fig:l_scan_Ir05}) 
reveals a period that is about 17\,\% larger, pointing towards a different 
origin of the corresponding excitation. It stems from Ir pairs 
where one Ir ion is located on a $2a$ site, see Sec.\ \ref{sec:2a}.

\begin{figure}[t]
\centering
\includegraphics[width=\columnwidth]{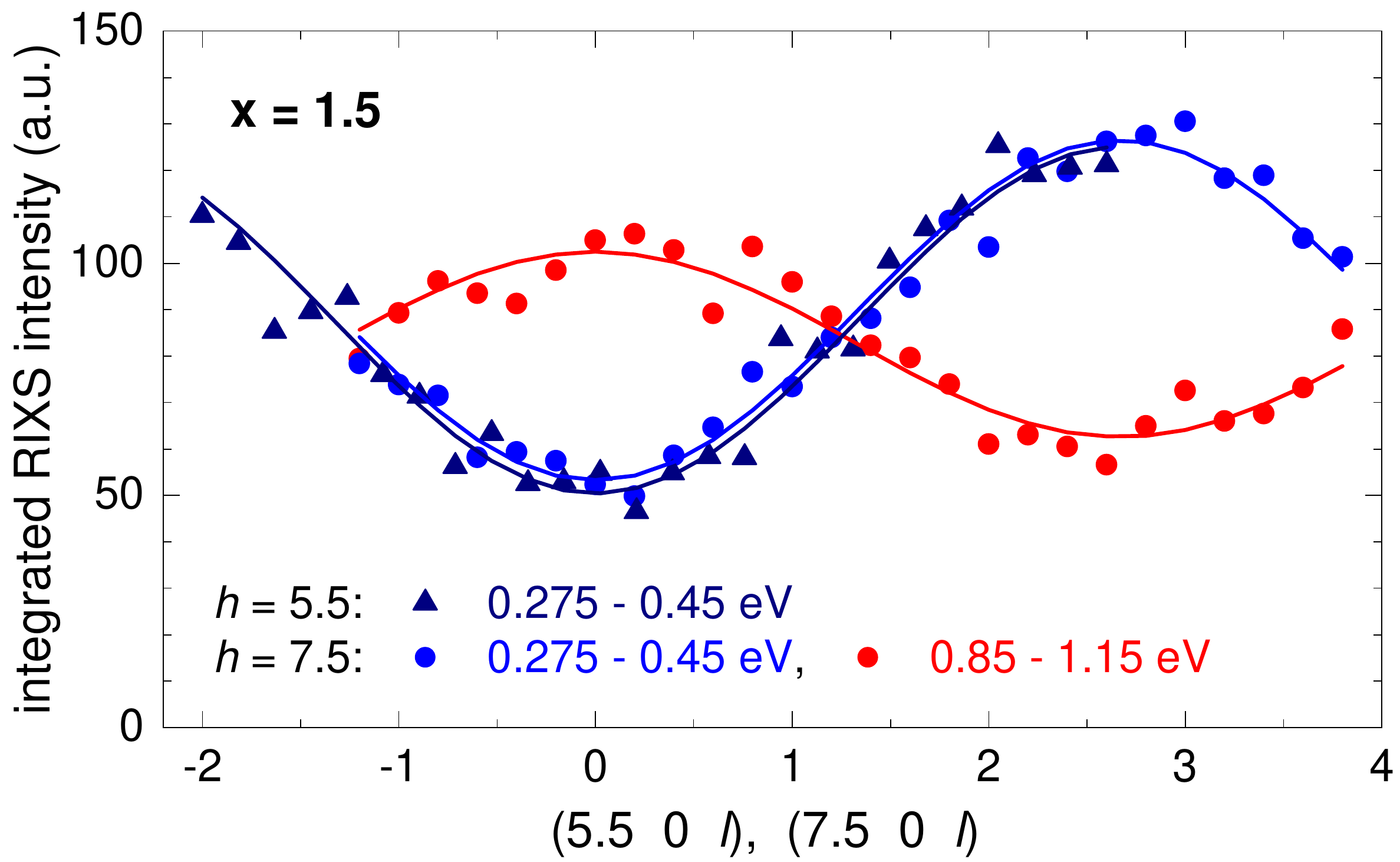}
\caption{RIXS interference patterns for $x$\,=\,1.5.
	The 0.95\,eV dimer peak shows $\cos^2(q_c\,d/2)$ behavior (red), 
	cf.\ Fig.\ \ref{fig:4samples}(c),  
	while a $\sin^2(q_c\,d/2)$ modulation is observed for integration from 
	0.275 to 0.45\,eV, in agreement with Fig.\ \ref{fig:l_scan_Ir05}(b). 
    Data were collected on a $(100)$ surface with $c$ (nearly) perpendicular to the 
    scattering plane, i.e., rotated by $90^\circ$ with respect to the geometry used 
    in Fig.\ \ref{fig:l_scan_Ir05}(b).
    A finite range of $l$ can be covered by tilting the sample. 
	Solid lines: fits according to $I(l) = a_0 \sin^2(\pi l /2Q_d) + c_0$ and 
	equivalently for the $\cos^2$ curve, using the period 
	$2Q_d$\,=\,$5.34 \times 2\pi/c$ determined from the data in 
	Fig.\ \ref{fig:l_scan_Ir05}(b).
}	
\label{fig:l_scan_perp}
\end{figure}

\begin{figure}[t]
\centering
\includegraphics[width=\columnwidth]{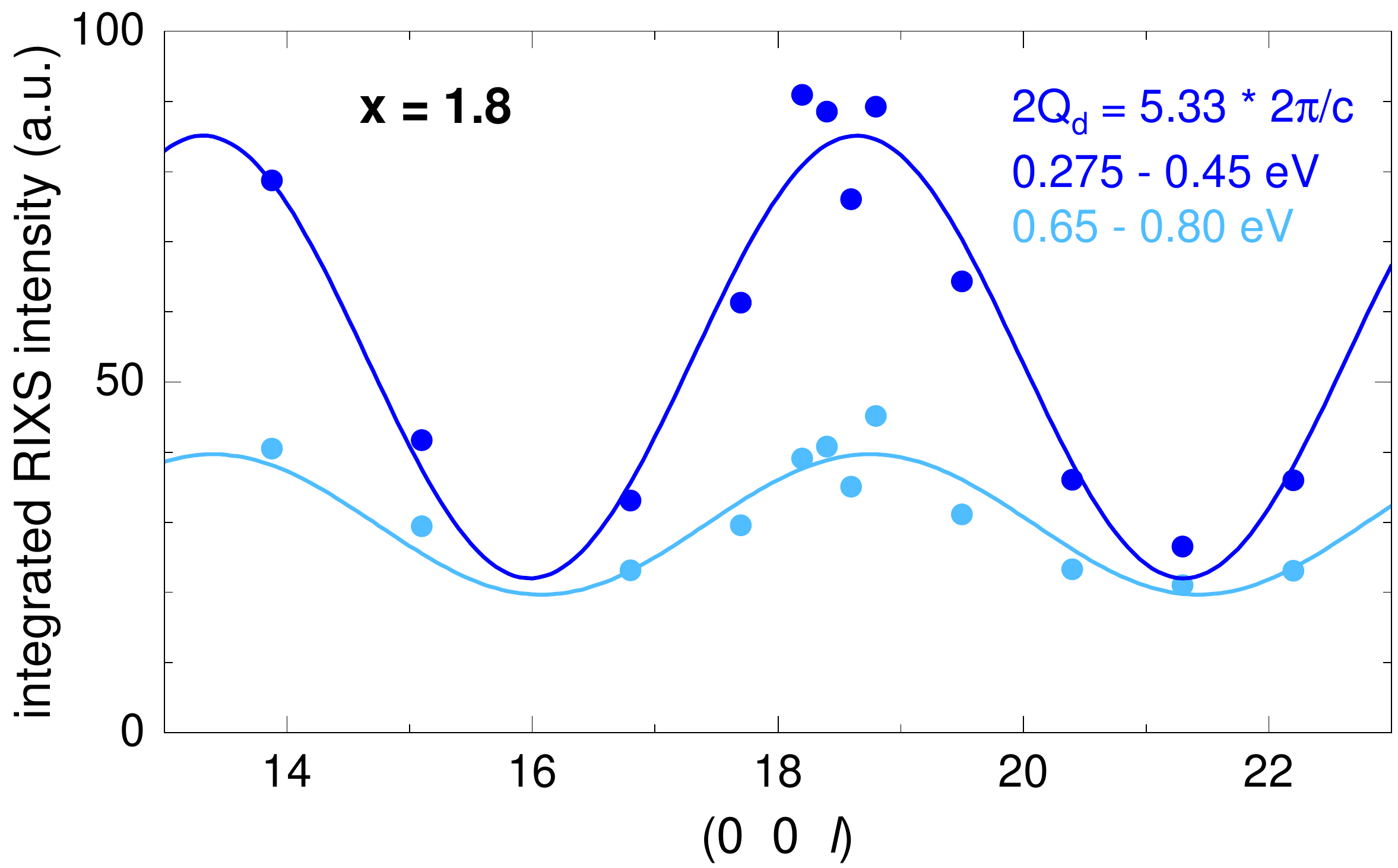}
\caption{Modulated RIXS intensity for $x$\,=\,1.8 for large $l$. 
	In contrast to the $\mathbf{q}$ scans with a fixed energy window
	depicted in Figs.\ \ref{fig:l_scan_Ir05} and \ref{fig:l_scan_perp}, 
	the data (symbols) were obtained from a series of RIXS spectra, 
	measured up to 1.5\,eV, by integrating the intensity of the dominant peak 
	from 0.275 to 0.45\,eV (blue) and from 0.65 to 0.80\,eV (light blue). 
	Particularly large values of $l$ are reached by measuring on a $(001)$ surface.
	Solid lines: Fits yield the period $2Q_d$\,=\,$(5.33 \pm 0.05)\times 2\pi/c$. 
	With the knowledge of the $\sin^2(q_cd/2)$ character from Fig.\ \ref{fig:l_scan_Ir05}, 
	the large $l$ value allows us to determine the period with high accuracy.
	}	
	\label{fig:l_scan_18}
\end{figure}

The sinusoidal intensity modulation described by Eq.\ (\ref{eq:Iqc}) was also observed 
in RIXS on magnetic excitations in honeycomb Na$_2$IrO$_3$ where dominant Kitaev 
interactions yield dynamical spin-spin correlations that are restricted to 
nearest neighbors on a bond, 
forming the dynamical magnetic equivalent of a dimer \cite{Revelli20}. 
A sinusoidal modulation in $\mathbf{q}$ space reflects the Fourier transform of a dimer.
In the case of RIXS, 
this intensity modulation reveals the dynamical structure factor of 
a dimer excitation. In other words, it is equivalent to an inelastic realization of a 
double-slit-type interference pattern \cite{Gelmukhanov94,Ma94,Ma95,Gelmukhanov21,Revelli19}. 
To calculate the corresponding total RIXS amplitude $A_f(\mathbf{q})$, one has to 
sum up all scattering processes that lead to a given final excited state $|f\rangle$. 
The RIXS process at the Ir $L_3$ edge proceeds via an intermediate state with a 
$2p$ core hole that is strongly localized on a given Ir site. In the case of Ir$_2$O$_9$ dimers 
with quasimolecular orbitals, the scattering process may occur on each of the two Ir sites 
$\mathbf{R}_i$ over which the holes are delocalized in both the ground state $|0\rangle$ 
and the excited state $|f\rangle$. The summation thus has to run over both sites, 
\begin{eqnarray}
A_f(\mathbf{q}) & \propto & \langle f| \sum_{i\in\{1,2\}} e^{i \mathbf{qR}_i}\, M_{\mathbf{R}_i}\, |0\rangle \, , 
\end{eqnarray}
where $M_{\mathbf{R}_i}$ denotes the dipole matrix element for the RIXS process via 
site $\mathbf{R}_i$ and $e^{i \mathbf{qR}_i}$ is the corresponding phase factor. 
Assuming a dimer with inversion symmetry, the matrix elements on the two sites share the 
same modulus but may differ in sign, which yields 
\begin{eqnarray}
A^{\rm inv}_f(\mathbf{q}) & \propto & e^{i q_c d/2} \pm e^{-i q_c d/2} \, .
\label{eq:exp}
\end{eqnarray}
Since the RIXS intensity is proportional to the amplitude squared, this yields either 
$\sin^2$ or $\cos^2$ behavior, depending on the parity of $|0\rangle$ and $|f\rangle$. 
However, a single Ir$_2$O$_9$ bi-octahedron exhibits mirror symmetry but no inversion 
symmetry, giving rise to a mixture of $\sin(q_c\, d/2)$ and $\cos(q_c\,d/2)$ terms 
in the amplitude. In Ba$_3$Ti$_{3-x}$Ir$_x$O$_9$, the bi-octahedra in adjacent layers 
are rotated by $\pi$ around $c$ with respect to each other, hence the two different 
dimer orientations are transformed into each other via inversion, see Fig.\ \ref{fig:structure}. 
Summing the intensities of the two orientations cancels the terms odd in $q_c$ 
such that only $\sin^2(q_c\,d/2)$ and $\cos^2(q_c\,d/2)$ terms survive \cite{Revelli19}. 
For the expected RIXS intensity this yields
\begin{equation}
I_f(q_c) \propto u(q_c) \sin^2(q_c \,d/2) + v(q_c) \cos^2(q_c \,d/2) \, , 
\label{eq:Iqc_mixed}
\end{equation}
where the $q_c$ dependences of the prefactors again mainly reflect polarization effects. 
For $u > v$, this explains the dominant $\sin^2$ behavior described by Eq.\ (\ref{eq:Iqc}) 
and observed in Figs.\ \ref{fig:l_scan_Ir05}, \ref{fig:l_scan_perp}, and 
\ref{fig:l_scan_18} for the peak at 0.35\,eV.\@ 
The opposite behavior with $u < v$ resulting in a dominant $\cos^2$ term 
is found for the 0.95\,eV peak, see Fig.\ \ref{fig:l_scan_perp}. As discussed below, 
this indicates a spin-flip character of the 0.95\,eV peak. 
The period $2Q_d$\,=\,$2\pi/d$ provides a measure of the intradimer Ir-Ir distance $d$. 
We find $2Q_d$\,=\,$(5.34\pm 0.04)\times 2\pi/c$ 
which is equivalent to $d$\,=\,$(2.66 \pm 0.02)$\,\AA\ at 20\,K, 
in agreement with the value 2.65\,\AA\ determined in elastic x-ray diffraction 
at 300\,K \cite{Sakamoto06}.

The two dimer sites are displaced along the $c$ axis. This explains the intensity 
modulation as a function of $l$ and predicts that dimer features are insensitive 
to $h$. This is indeed observed above 0.2\,eV, see Figs.\ \ref{fig:spectra_h} 
and \ref{fig:h_scan}. 
In contrast, the RIXS peak at 0.15\,eV exhibits a pronounced $h$ dependence 
as well as a larger period as a function of $l$, see Fig.\ \ref{fig:l_scan_Ir05}(b), 
which points to a different microscopic origin, see Sec.\ \ref{sec:2a}.

A quasimolecular dimer character has been demonstrated in RIXS measurements 
on the sister compounds Ba$_3$CeIr$_2$O$_9$ and Ba$_3$InIr$_2$O$_9$ with two 
and three holes per dimer, respectively \cite{Revelli19,Revelli22}. 
The short intradimer Ir-Ir distance of about 2.5\,-\,2.6\,\AA\ \cite{Doi04} 
gives rise to a very large hopping of the order of 0.5\,-\,1\,eV \cite{Revelli19,Revelli22} 
and a corresponding large splitting of bonding and antibonding quasimolecular 
orbitals in which the holes are fully delocalized over a given dimer.
As in Ba$_3$CeIr$_2$O$_9$ with Ce$^{4+}$ ions, the Ir$_2$O$_9$ dimers in 
Ba$_3$Ti$_{3-x}$Ir$_x$O$_9$ host two holes that can be placed in the lowest 
quasimolecular orbital, giving rise to a nonmagnetic ground state \cite{Revelli19}. 
The RIXS data of Ba$_3$CeIr$_2$O$_9$ show three features peaking at about 
0.7, 1.0, and 1.2\,eV.\@ Similarly, the dimer features in Ba$_3$Ti$_{3-x}$Ir$_{x}$O$_9$
for large $x$ can be roughly described by three peaks at about 0.35, 0.7, and 0.95\,eV, 
suggesting a reduced energy scale for $M$\,=\,Ti compared to Ce. 
This tentatively can be ascribed to the larger Ir-Ir distance $d$, which amounts 
to 2.65\,\AA\ \cite{Sakamoto06} in the Ti compounds and 2.54\,\AA\ \cite{Revelli19} 
in Ba$_{3}$CeIr$_{2}$O$_{9}$. In a simple picture, a larger Ir-Ir distance $d$ 
yields reduced hopping and hence a smaller bonding-antibonding splitting. 
The common character of the dimer states in Ba$_3$CeIr$_2$O$_9$ and 
Ba$_3$Ti$_{3-x}$Ir$_x$O$_9$ is supported by the similar $\mathbf{q}$ dependences. 
In Ba$_3$CeIr$_2$O$_9$, the two lower-energy dimer features show a pronounced 
$\sin^2(q_cd/2)$ behavior while the third peak at 1.2\,eV exhibits 
$\cos^2(q_cd/2)$ behavior which is strongly suppressed at large $l$ \cite{Revelli19}. 
The same can be observed in Ba$_3$Ti$_{3-x}$Ir$_x$O$_9$. 
The two features at 0.35 and 0.7\,eV show a $\sin^2(q_c d/2)$ intensity modulation, 
see Figs.\ \ref{fig:l_scan_Ir05} and \ref{fig:l_scan_18}, 
while the third peak at 0.95\,eV hardly depends 
on $l$ for large $l$, cf.\ Fig.\ \ref{fig:4samples}(d), but indeed shows 
$\cos^2(q_c d/2)$ behavior for small $l$, 
see Figs.\ \ref{fig:l_scan_perp} and \ref{fig:4samples}(c).
For $M$\,=\,Ce, the distinct behavior of the third peak has been shown to arise 
from the spin-flip character of this excitation \cite{Revelli19}. 
Altogether this suggests that the peak at 0.95\,eV for $x$\,=\,1.5 and 1.8 
corresponds to a singlet-to-triplet dimer excitation.

In contrast to the related compounds Ba$_3$$M$Ir$_2$O$_9$, the physics of 
Ba$_3$Ti$_{3-x}$Ir$_x$O$_9$ is governed by strong Ti-Ir site disorder. 
This is highlighted by the coexistence of Ir$_2$O$_9$ dimers and single Ir sites 
in TiIrO$_9$ units for not too small values of $x$. 
If dimer formation was energetically unfavorable, it could be avoided for $x$\,=\,0.5, 
and the same argument holds for single sites at large $x$. 
The coexistence thus suggests that neither dimers nor single sites are strongly preferred, 
pointing towards a random distribution of Ir and Ti ions over the dimer sites, 
in agreement with previous results \cite{Dey12,Kumar16,Lee17,Sakamoto06}. 
Combined with the nonmagnetic character of the dimer ground state, this disorder 
provides the key to understand the surprising $x$ dependence of the magnetic susceptibility, 
see Fig.\ \ref{fig:chi}. 
For small $x$, the susceptibility reflects the dominant contribution of dilute 
$j$\,=\,1/2 moments located on individual Ir sites. With increasing $x$, however, 
the relative contribution of nonmagnetic dimers is enhanced, which may even lead to 
a suppression of $\chi(T)$ upon strongly increasing the Ir content.

\subsection{Magnetic excitation on 2a sites}
\label{sec:2a}

\begin{figure}[t]
\centering
\includegraphics[width=\columnwidth]{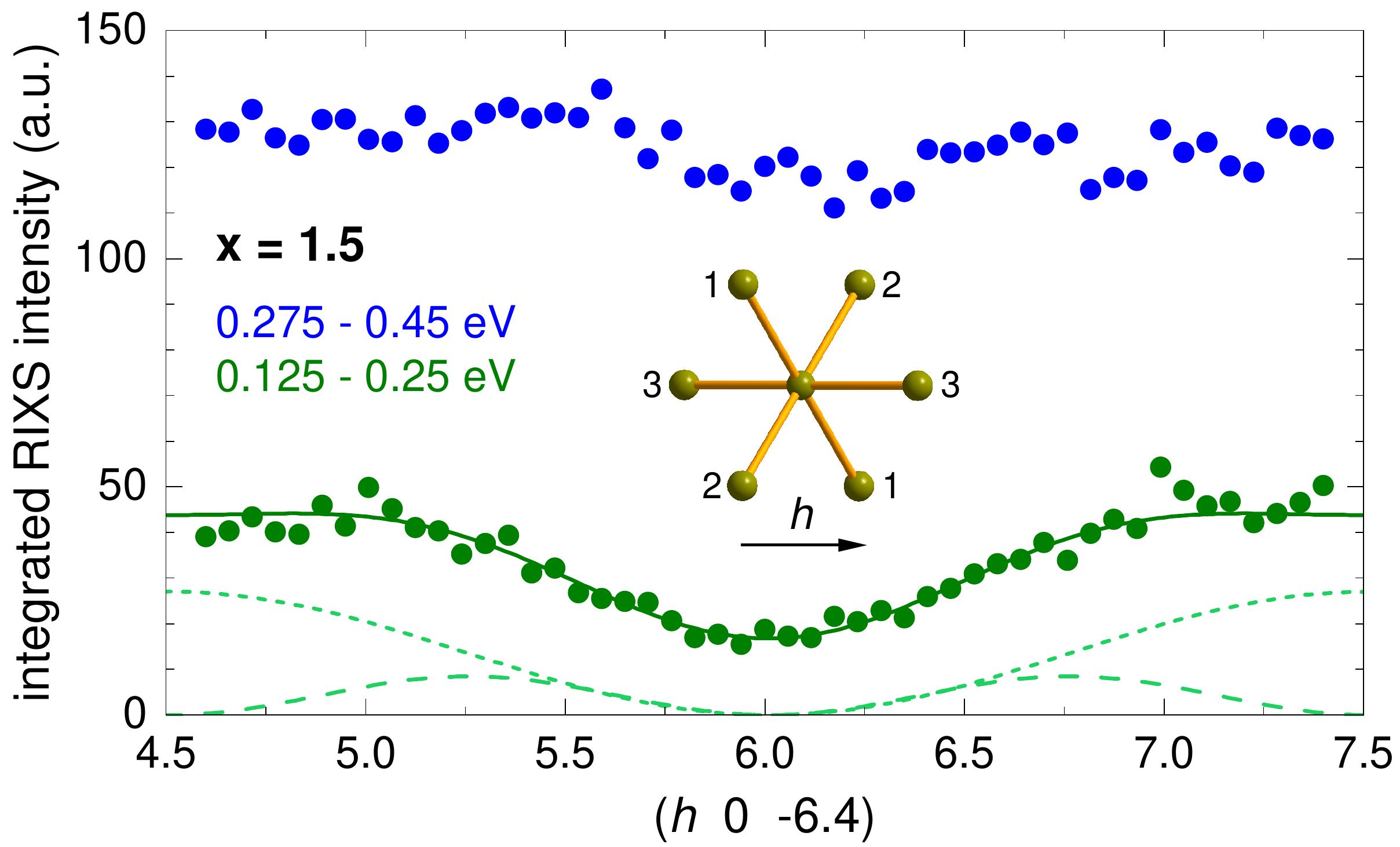}
\caption{RIXS interference patterns of Ba$_{3}$Ti$_{1.5}$Ir$_{1.5}$O$_{9}$ as a 
	function of $h$. The dominant dimer peak at 0.35\,eV, integrated from 0.275\,eV 
	to 0.45\,eV (blue), does not depend on $h$ since the dimer axis is parallel to $c$, 
	cf.\ Eq.\ (\ref{eq:Iqc}) and Fig.\ \ref{fig:spectra_h}. 
	In contrast, the 0.15\,eV peak (dark green) shows a modulation with two different 
	periods along $h$. The solid line depicts the total fit, cf.\ Eq.\ (\ref{eq:fit}), 
	while the light green lines show the two contributions with periods 3 and 3/2 in $h$. 
	Accordingly, the 0.15\,eV peak corresponds to a magnetic excitation for a pair of 
	Ir ions located on neighboring $2a$ and $4f$ sites. 
	Sketch: projection of the Ir$^{2a}$-O-Ir$^{4f}$ bonds onto the $ab$ plane, 
	illustrating the two values of $|\mathbf{q}\cdot\mathbf{r}_i|$ for 
	$\mathbf{q}$\,=\,($h$\,\,0\,\,0)\,r.l.u., cf.\ Eqs.\ (\ref{eq:r1})-(\ref{eq:r3}). 
	The numbers $i\in\{1,2,3\}$ denote the six bonds $\pm \mathbf{r}_i$ for 
    a given $2a$ site. The Ir$^{2a}$O$_6$ octahedra in the neighboring layers are rotated 
    by $\pi$ around $c$, giving rise to the identical projection onto the $ab$ plane.
}
	\label{fig:h_scan}
\end{figure}

In the following, we address the distinct properties of the 0.15\,eV peak, 
see Figs.\ \ref{fig:spectra_h},  \ref{fig:l_scan_Ir05}(b), and \ref{fig:h_scan}, 
and demonstrate how RIXS interferometry allows us to unravel its microscopic origin. 
For $x$\,=\,1.5, integration of the RIXS intensity as a function of $q_c$ 
in the energy range 0.125 to 0.25\,eV yields a modulation period 
$2Q_{2a}$\,=\,$(6.27 \pm 0.08)\times 2\pi/c$ 
that is about 17\,\% larger than the dimer value $2Q_d$, see Fig.\ \ref{fig:l_scan_Ir05}(b). 
The sinusoidal pattern again points towards the interference of scattering on two sites, 
but their distance projected onto the $c$ axis amounts to d$_{2a}$\,=\,$(2.26 \pm 0.03)$\,\AA. 
This agrees with the difference of the $z$ components of $2a$ and $4f$ sites of 2.22\,\AA\ 
measured in x-ray diffraction at 300\,K \cite{Sakamoto06}, pointing towards the presence of 
Ir ions on $M$ $(2a)$ sites interacting  with Ir ions on $4f$ sites. 
These two sites are also displaced perpendicular to $c$, cf.\ Fig.\ \ref{fig:structure}, 
giving rise to the $h$ dependence of the 0.15\,eV feature shown in 
Figs.\ \ref{fig:spectra_h} and \ref{fig:h_scan}.

Taking a $2a$ site as the origin of our coordinate frame, it has six $4f$ neighbors 
at $\pm \mathbf{r}_i$ with $i \in \{1,2,3\}$, see right part of Fig.\ \ref{fig:structure} 
and sketch in Fig.\ \ref{fig:h_scan}. 
For simplicity, we consider separate pairs that are built from the Ir ion on the $2a$ site 
and by \textit{one} Ir neighbor on one of the six neighboring $4f$ sites. 
This can be motivated by the fact that the 0.15\,eV peak is only 
observed for large $x$, in which case most of the neighboring bi-octahedra will be occupied 
by two Ir ions with two holes forming a stable singlet in a quasimolecular orbital. 
We neglect the interaction with such singlets, and we may also neglect those bi-octahedra 
that are occupied by two Ti ions. Relevant to us are the TiIrO$_9$ bi-octahedra where 
the Ir $j$\,=\,1/2 moment on a $4f$ site may interact with one on a $2a$ site, 
as indicated by the thick black line on the right side of Fig.\ \ref{fig:structure}. 
For such pairs, we expect a sinusoidal intensity modulation $\sin^2(\mathbf{q\cdot r}_i/2)$, 
equivalent to Eq.\ (\ref{eq:Iqc}). We find
\begin{eqnarray}
\label{eq:r1}
\frac{1}{2}\,\mathbf{q} \cdot \mathbf{r}_{1,4} & = & \pm \left[ (h+2k) \frac{\pi}{3} 
\pm l\, \frac{\pi}{6.27} \right] \, , \\
\label{eq:r2}
\frac{1}{2}\,\mathbf{q} \cdot \mathbf{r}_{2,5} & = & \pm \left[ (h-k) \frac{\pi}{3} 
\pm l\,\frac{\pi}{6.27} \right] \, , \\
\label{eq:r3}
\frac{1}{2}\,\mathbf{q} \cdot \mathbf{r}_{3,6} & = & \pm \left[ -(2h+k) \frac{\pi}{3} 
\pm l\,\frac{\pi}{6.27} \right] \, 
\end{eqnarray}
for $2Q_{2a}$\,=\,$6.27 \times 2\pi/c$.  
The indices $i\in\{1,2,3\}$ and $i\in\{4,5,6\}$ of the bonds $\pm \mathbf{r}_i$ 
refer to $2a$ sites in adjacent layers. The corresponding Ir$^{2a}$O$_6$ octahedra 
are rotated by $\pi$ around $c$, giving rise to the different signs in front of the $l$ term.
The overall sign of $\mathbf{r}_i$ is irrelevant for the $\sin^2(\mathbf{q\cdot r}_i/2)$ term.
Note that all possible pairs yield the same $l$ dependence for 
$h$\,=\,$k$\,=\,0. This agrees with the sinusoidal modulation with a single period $2Q_{2a}$ 
observed for the green symbols in Fig.\ \ref{fig:l_scan_Ir05}(b), 
which in turn strongly supports the simple picture of separate pairs. 
The data in Fig.\ \ref{fig:l_scan_Ir05}(b) were measured with $h$\,=\,6.1, 
which is nearly equivalent to $h$\,=\,0.

Treating each pair separately, we add the intensities for all six kinds of pairs described 
by Eqs.\ (\ref{eq:r1}) -- (\ref{eq:r3}). This summation over the two different layers of 
Ir$^{2a}$O$_6$ octahedra, rotated by $\pi$ around $c$, yields an even function of $l$, 
as mentioned above for the dimers,
\begin{eqnarray}
\label{eq:hkl}
I[(h\,\,k\,\,l)] & \propto & \sum_{i=1\,{\rm to}\,6} \sin^2(\mathbf{q} \cdot \mathbf{r}_{i}/2) 
\\
	\nonumber
	& = &
	3 + \cos\left( l\,\frac{2\pi}{6.27} \right) \cdot 
	\bigg[2\sin^2\left((h+2k)\frac{\pi}{3} \right)  
	\\
\nonumber
	& + &   2\sin^2\left((h-k)\frac{\pi}{3} \right) 
	      + 2\sin^2\left((2h+k)\frac{\pi}{3} \right) -3 
	         \bigg]  \, .
\end{eqnarray}
The projection of the different $2a$-$4f$ pairs onto the $ab$ plane, 
sketched in Fig.\ \ref{fig:h_scan}, yields a characteristic dependence on $h$ and $k$  
for constant $l$ that allows us to identify corresponding pair excitations.
Figure \ref{fig:h_scan} shows the integrated RIXS intensity of the 0.15\,eV peak 
as a function of $h$ for $k$\,=\,0 and $l$\,=\,$-6.4$. The latter is nearly equivalent 
to $l \approx 0$ since it is close to a full period $2Q_{2a}$\,=\,$6.27 \times 2\pi/c$.
For this $h$ scan, the model predicts two periods in $h$, namely 3 and 3/2,
\begin{equation}
I[(h\,\,0\,\,0)]  \propto 2 \sin^2\left(h \frac{\pi}{3}\right)+ \sin^2\left(2h \frac{\pi}{3}\right) \, , 
\label{eq:h00}
\end{equation}
due to the two different projections of the $\mathbf{r}_i$ on ($h$\,\,0\,\,0).
The data in Fig.\ \ref{fig:h_scan} are well described by a fit (dark green line) based on
\begin{equation}
I_{\rm fit}(h) = a_0 \left[ 2 \sin^2\left(h \frac{\pi}{3}\right) + 
b_0 \sin^2\left(2h \frac{\pi}{3}\right) \right] + c_0 \, ,
\label{eq:fit}
\end{equation}
where the two contributions are depicted by the light green lines. 
The observation of the two periods strongly supports our simple model, even though 
we find a reduced prefactor $b_0$\,=\,0.63 for the second term compared to 
Eq.\ (\ref{eq:h00}).
This scenario of separate pairs of Ir ions on $2a$ and $4f$ sites also explains 
the intriguing $h$ dependence of the 0.15\,eV peak revealed in Fig.\ \ref{fig:spectra_h}. 
For $l$\,=\,16.8, the model indeed predicts a maximum of intensity for $h$\,=\,0 
and a minimum for $h$\,=\,2, while the opposite is expected for $l$\,=\,19.6, 
as observed experimentally.

The excitation energy of 0.15\,eV roughly can be motivated via the bonding geometry. 
The IrO$_6$ octahedra around the $2a$ and $4f$ sites share a common corner, giving rise to 
a 180$^\circ$ Ir-O-Ir bond, see Fig.\ \ref{fig:structure}. 
This bonding geometry is equivalent to the case of square-lattice Sr$_{2}$IrO$_{4}$ 
with strong Heisenberg exchange $J$ and magnon energies extending up to roughly 
0.2\,eV \cite{Kim12,KimNatComm14}.
For a single bond of two moments coupled by $J$, the excitation energy is of the 
same order of magnitude as found for zone-boundary magnons on the square lattice, 
which roughly explains the peak energy of 0.15\,eV.\@ 
This scenario of exchange-coupled local moments in contrast to quasimolecular orbitals 
is based on the fact that hopping $t$ is significantly smaller in corner-sharing 
geometry than for face-sharing octahedra. 
The ratio of the on-site Coulomb repulsion $U$ over $t$ is then large enough 
to suppress a quasimolecular character and the physics is described 
by two $j$\,=\,1/2 moments coupled by an exchange interaction $J \propto t^2/U$.
Note that a similar RIXS feature was observed at 0.23\,eV in Ba$_3$InIr$_2$O$_9$ 
with three holes per dimer and magnetic $j_{\rm dim}$\,=\,3/2 dimer moments \cite{Revelli22}. 
There, about 7\,\% of Ir ions were observed on the In $2a$ sites.

Altogether, our results provide strong evidence for the identification of the 
0.15\,eV peak as a magnetic excitation of a pair of exchange-coupled Ir 
$j$\,=\,1/2 moments located on neighboring $2a$ and $4f$ sites. 
This assignment is further supported by the observed symmetry, i.e., the dominant 
$\sin^2(\mathbf{q \cdot r}_i/2)$ character of the interference pattern plotted 
in Figs.\ \ref{fig:l_scan_Ir05}(b) and \ref{fig:h_scan}, cf.\ Eq.\ (\ref{eq:fit}).
The corner-sharing octahedra of a $2a$-$4f$ pair show the same orientation, 
see Fig.\ \ref{fig:structure}. In this case we expect the minus sign in 
Eq.\ (\ref{eq:exp}) for a spin-flip excitation, in agreement with the observed 
$\sin^2(\mathbf{q \cdot r}_i/2)$ behavior. 
In contrast, the two IrO$_6$ octahedra of a face-sharing dimer are rotated by $\pi$ 
around $c$ with respect to each other. 
For the intradimer spin-flip at 0.95\,eV, this contributes a further minus sign
and yields the $\cos^2(q_c\,d/2)$ behavior observed in Fig.\ \ref{fig:l_scan_perp}.

\section{Conclusions}

The physics of Ba$_3$Ti$_{3-x}$Ir$_{x}$O$_9$ is dominated by strong disorder. 
The similar radii and equal valence of Ir$^{4+}$ and Ti$^{4+}$ ions yield mixed crystals with 
pronounced Ir-Ti site mixing.  For $x\geq 0.5$, our RIXS spectra of samples with 
$x$\,$\in$\,$\{0.3, 0.5, 1.5, 1.8\}$ demonstrate the coexistence of single-site $j$\,=\,1/2 
moments and Ir$_2$O$_9$ dimers, as expected for a random distribution of Ir and Ti ions. 
Individual spin-orbit-entangled $j$\,=\,$1/2$ moments prevail for small Ir content $x$ 
while the density of Ir$_2$O$_9$ dimers increases with $x$. 
Due to the large intradimer hopping, the dimers host quasimolecular orbitals in which 
the holes are fully delocalized over the two dimer sites. Both holes occupy the lowest 
binding orbital in the dimer ground state, forming a nonmagnetic $j_{\rm dim}$\,=\,0 singlet. 
The coexistence of nonmagnetic dimers and $j$\,=\,1/2 moments explains the 
at first sight unconventional behavior of the magnetic susceptibility which is very similar 
for $x$\,=\,0.5 and 1.5, i.e., for two samples in which the concentration of magnetic 
Ir ions differs by a factor of three.

These compounds offer a remarkable example for the unusual role that disorder may play in 
a cluster Mott insulator. Substituting a magnetic ion by a nonmagnetic one typically yields 
a vacancy in the magnetic system. In a cluster, however, it changes the character of the 
magnetic moment. In Ba$_3$Ti$_{3-x}$Ir$_{x}$O$_9$, cation disorder gives rise to different 
kinds of magnetic moments, and the different moments experience very different coupling 
strengths. On a dimer, the energy scale for singlet-to-triplet excitations is given by 
the RIXS peak at 0.95\,eV, reflecting the large intradimer hopping. In contrast, the large 
spatial separation between dimers strongly suppresses interdimer interactions \cite{Bhowal19}. 
Considering, e.g., $j$\,=\,1/2 moments on neighboring IrTiO$_9$ bi-octahedra, 
we expect exchange interactions on the order of meV, i.e., three orders of magnitude smaller 
than the intradimer singlet-to-triplet excitation energy within Ir$_2$O$_9$ units. 
Additionally, we observe pairs of exchange-coupled $j$\,=\,1/2 moments that arise from Ir ions 
located on adjacent $2a$ and $4f$ sites. These are connected via a $180^\circ$ bond that 
yields a large Heisenberg exchange coupling $J$ similar to the case of square-lattice 
Sr$_2$IrO$_4$, giving rise to a magnetic excitation at $0.15$\,eV.\@

RIXS interferometry enabled us to unravel this complex disorder scenario. 
The RIXS spectra and in particular the $\mathbf{q}$ dependence of the RIXS intensity provide 
clear fingerprints of the different types of moments. The excitations of local single-site 
$j$\,=\,1/2 moments are independent of $\mathbf{q}$. In contrast, a sinusoidal modulation 
of the RIXS intensity of orbital excitations is a characteristic hallmark of 
quasimolecular dimer orbitals, and a predominant 
$\sin^2(\mathbf{q\cdot r}/2)$ or $\cos^2(\mathbf{q\cdot r}/2)$ behavior typifies the 
symmetry and character of the excited states. The RIXS studies of the sister compounds 
Ba$_3$CeIr$_2$O$_9$ and Ba$_3$InIr$_2$O$_9$ \cite{Revelli19,Revelli22} were focused 
on establishing the distinct quasimolecular character for two or three holes per dimer. 
In Ba$_3$Ti$_{3-x}$Ir$_{x}$O$_9$, the periodicity of the modulation allows us to identify 
which \textit{sites} contribute to a given excitation. We can distinguish Ir$_2$O$_9$ dimers 
from $2a$-$4f$ pairs via the distinct modulation period along $l$ as well as via the 
absence or presence of a characteristic modulation along $h$ or $k$. 
We expect that our results will trigger many further experimental investigations to exploit 
the stunning power of RIXS interferometry to study the electronic structure of dimers, 
trimers, and larger clusters, even in the presence of disorder.

\acknowledgments 

We thank A. Revelli for experimental support and useful discussions. 
We gratefully acknowledge the European Synchrotron Radiation Facility for providing 
beam time and technical support. 
Furthermore, we acknowledge funding from the Deutsche Forschungsgemeinschaft 
(DFG, German Research Foundation) through Project No.\ 277146847 -- CRC 1238 
(projects A02 and B03) and 
Project No.\ 247310070 -- CRC 1143 (project A05). 
M.H.\@ acknowledges partial funding by the Knut and Alice Wallenberg Foundation 
as part of the Wallenberg Academy Fellows project and by the Swedish Research Council 
through grant no.\ 2017.0157.

\end{document}